\documentclass[twocolumn,showpacs,preprintnumbers,amsmath,amssymb]{revtex4}

\usepackage{graphicx}
\usepackage{dcolumn}
\usepackage{bm}

\newcommand{\f}[2]{\frac{#1}{#2}}

\begin{document}

\preprint{UAB-FT-605}

\title{Compatibility of CAST search with axion-like interpretation of PVLAS results}
\author{Eduard Mass{\'o}}
\author{Javier Redondo}%
\affiliation{%
Grup de F{\'i}sica Te{\`o}rica and IFAE, Universitat Aut{\`o}noma de Barcelona,\\
 E-08193 Bellaterra (Barcelona), Spain.}

\date{\today}

\begin{abstract}
The PVLAS collaboration has results that may be interpreted in terms of
a light axion-like particle, while the CAST collaboration has not found any
signal of such particles. We propose
a particle physics model with paraphotons and with a low energy scale
in which this apparent inconsistency is circumvented.
\end{abstract}

\maketitle
%
Very recently, the PVLAS collaboration has announced the observation
of a rotation of the plane of polarization of laser light
propagating in a magnetic field \cite{Zavattini:2005tm}. This
dichroism of vacuum in magnetic fields may be explained as the
oscillation of photons into very light particles $\phi$. If true,
this would be of course a revolutionary finding \cite{nature}.

The lagrangian that would describe the necessary $\phi\gamma\gamma$
coupling is
\begin{equation}
{\cal L}_{\phi\gamma\gamma}=\frac{1}{8M}\,
\epsilon_{\mu\nu\rho\sigma}F^{\mu\nu} {F}^{\rho\sigma}\, \phi
\label{L}
\end{equation}
when $\phi$ is a pseudoscalar, and when it is a scalar is
\begin{equation}
{\cal L}_{\phi\gamma\gamma}=\frac{1}{4M}\, F^{\mu\nu}F_{\mu\nu}\,
\phi \label{LS}
\end{equation}
with $F^{\mu\nu}$ the electromagnetic field tensor. We shall refer
to $\phi$ in both cases as an axion-like particle (ALP). Let us
remark that a transition to a spin-two particle contributes to the
polarization rotation negligibly \cite{Biggio:2006im}.

Either (\ref{L})  or (\ref{LS}) lead to $\gamma-\phi$ mixing in a
magnetic field and, if $\phi$ is light enough, to coherent
transitions that enhance the signal \cite{Maiani:1986md}.
Interpreted in these terms, the PVLAS observation
\cite{Zavattini:2005tm} leads to a mass for the ALP
\begin{equation}
1\ {\rm meV} \lesssim m_\phi \lesssim 1.5\ {\rm meV} \label{m_phi}
\end{equation}
and to a coupling strength corresponding to
\begin{equation}
2 \times 10^5\ {\rm GeV} \lesssim M \lesssim 6\times 10^5\ {\rm GeV}
\ \ \ . \label{M_PVLAS}
\end{equation}
Of course we would like to have an independent test of such an
interpretation. There are ongoing projects that will in the near
future probe $\gamma-\phi$ transitions \cite{ringwald}. In the
meanwhile we should face the problem of the apparent inconsistency
between the value (\ref{M_PVLAS}) and other independent results,
namely, the CAST observations \cite{Zioutas:2004hi} on the one hand,
and the astrophysical bounds on the coupling of ALPs to photons on
the other hand \cite{Raffelt:1996wa}.

The CAST collaboration has recently published  \cite{Zioutas:2004hi}
a limit on the strength of (\ref{L}) or (\ref{LS}). A light particle
coupled to two photons would be produced by Primakoff-like processes
in the solar core. CAST is an helioscope \cite{Sikivie:1983ip} that
tries to detect the $\phi$ flux coming from  the Sun,  by way of the
coherent transition of $\phi$'s to X-rays in a magnetic field. As no
signal is observed they set the bound
\begin{equation}
M > 0.87 \times 10^{10}\ {\rm GeV}  \ \ \ , \label{M_CAST}
\end{equation}
which is in strong disagreement with (\ref{M_PVLAS}).

Also, the production of $\phi$'s  in stars is constrained because
too much energy loss in exotic channels would lead to drastic
changes in the timescales of stellar evolution. Empirical
observations of globular clusters place a bound
\cite{Raffelt:1996wa}, again in contradiction with  (\ref{M_PVLAS}),
\begin{equation}
M > 1.7  \times 10^{10}\  {\rm GeV} \label{M_star}  \ \ \ .
\end{equation}

As it has been stressed in \cite{Masso:2005ym}, once we are able to
relax  (\ref{M_star}) we could also evade (\ref{M_CAST}). Indeed,
the CAST bound  assumes standard solar emission. From the moment we
alter the standard scenario we should revise (\ref{M_CAST}). In
\cite{Masso:2005ym,Jain:2005nh,Jaeckel:2006id} two ideas on how to
evade the astrophysical bound (\ref{M_star}) are presented. One
possibility is that the produced ALPs diffuse in the stellar medium,
so that  they are emitted with much less energy than originally
produced \cite{Masso:2005ym}. A second possibility is that the
production of ALPs is much less than expected because there is a
mechanism of suppression that acts in the stellar conditions. We
will present in this letter a paraphoton model with
 a low energy scale where the particle production in stars is
suppressed enough to accommodate both the CAST and the PVLAS
results.
\section{Triangle diagram and epsilon-charged particles \label{sectiontriangle}}
The physical idea beyond this letter is that to understand PVLAS and
CAST in an ALP framework we have to add some new physics structure
to the vertices (\ref{L}),(\ref{LS}). The scale of the new physics
should be much less than O(keV), the typical temperature in
astrophysical environments.

\begin{figure}[b]
  \vspace{2cm}
  \includegraphics{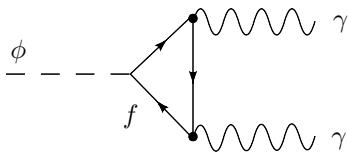}
  \caption{\it
    Triangle diagram for the $\phi\gamma\gamma$ vertex.
   \label{fig1} }
\end{figure}

We will assume that this structure is a simple loop where a new
fermion $f$ circulates; see Fig.(\ref{fig1}). The amplitude of the
$\phi\gamma\gamma$ diagram can be easily calculated and identified
with the coefficient in (\ref{L}) or (\ref{LS})
\begin{equation}
\frac{1}{M} = \frac{\alpha}{\pi} \frac{q_f^2}{v} \label{triangle}
\end{equation}
Here $\alpha=e^2/4\pi$, and the charge of the fermion $f$ is $eq_f$. The value of the mass-scale
$v$ depends on the $\phi\bar{f}f$ vertex. If $\phi$ is a pseudoscalar $v_{\rm PS}=m_f/g_{\rm PS}$
while if is scalar $v_{\rm S}=f(m_f,m_\phi)$ not far from $v_{\rm S}\sim m_f \sim m_\phi$ if
$m_f\sim m_\phi$. Finally if $\phi$ is a Goldstone boson $v_{\rm GB}$ is related to the scale of
breaking of the related global symmetry.

From (\ref{triangle}) we see that $M$,  the high energy scale (\ref{M_PVLAS}), is connected to $v$.
As we need $v$  to be a low energy scale,  $q_f$ should be quite small.

Paraphoton models \cite{Holdom:1985ag} naturally incorporate
 small charges. These models are QED extensions with extra
$U(1)$ gauge bosons. A small  mixing among the kinetic terms of the
gauge bosons leads to the exciting possibility that paracharged
exotic particles end up with a small induced electric charge
\cite{Holdom:1985ag}.

Getting a small charge for $f$ is not enough for our purpose since we need also production
suppression of exotic particles in stellar plasmas. With this objective, we will present a model
containing two paraphotons; if we allow for one of the paraphotons to have a mass, we will see we
can evade the astrophysical constraints and consequently the model will be able to accommodate all
experimental results. We describe it in what follows.
\section{A model with two paraphotons}
Let us start with the photon part of the QED lagrangian,
\begin{equation}
{\cal L}_0 =-\frac{1}{4}¼ F_0^{\mu\nu}F_{0\mu\nu}+ e\ j_{0\mu}
A_0^\mu
\end{equation}
where $j_{0\mu}$ is the electromagnetic current involving electrons, etc., $j_{0\mu} \sim
\bar{e}\gamma_\mu e + ...$. From the $U_0(1)$ gauge symmetry group, we give the step of assuming
$U_0(1)\times U_1(1)\times U_2(1)$ as the gauge symmetry group, with the corresponding gauge fields
$A_0$, $A_1$, and $A_2$. With all generality there will be off-diagonal kinetic terms in the
lagrangian, like $\epsilon_{01}\, F_0 F_1$ and $\epsilon_{02} F_0 F_2$ (Lorenz index contraction is
understood). We expect these mixings to be small if we follow the idea in \cite{Holdom:1985ag} that
ultramassive particles with 0,1,2 charges running in loops are the responsibles for them. We will
assume that these heavy particles are degenerate in mass and have identical 1 and 2 charges so that
they induce identical mixings $\epsilon_{01}=\epsilon_{02}\equiv \epsilon$.

To write the complete lagrangian we use  the matrix notation $A
\equiv (A_0, A_1, A_2)^T$ and $F \equiv (F_0, F_1, F_2)^T$,
\begin{equation}
{\cal L} =-\frac{1}{4}\, F^T {\cal M}_F\,  F+ \frac{1}{2}\, A^T
{\cal M}_A\,  A + e\sum_i\, j_{i} A_i \label{L_complete}
\end{equation}
We call $A_0$, $A_1$, and $A_2$ interaction fields because the interaction term in
\eqref{L_complete} is diagonal, i.e. the interaction photon is defined to couple directly only to
standard model particles. Here the kinetic matrix contains the mixings,
\begin{equation}
 {\cal M}_F  = \left( \begin{array}{ccc}
                                            1 & \epsilon & \epsilon  \\
                                            \epsilon & 1 &   0          \\
                                            \epsilon & 0 & 1
 \end{array}\right)
 \label{MF}
\end{equation}
In general the diagonal terms are renormalized, $1\rightarrow 1 + \delta$, and  there are terms
${\cal M}_{F12}$. However, they  do not play any relevant role here and we omit them.

As said, we need one of the paraphotons to be massive but it will prove convenient to work with a
general ${\cal M}_A  = {\rm Diag}\, \{ m_0^2, m_1^2, m_2^2 \}$. Also, in the last term of
(\ref{L_complete}) we see the currents $j_1$ and $j_2$ containing the paracharged exotic particles.
To reduce the number of parameters we have set the unit paracharge equal to the unit of electric
charge, so that there is a common factor $e$.

Diagonalization  involves first a non-unitary reabsorption of the $\epsilon$ terms in (\ref{MF}) to
have the kinetic part in the lagrangian in the canonical  form, $(-1/4) F^T F$. After this, we
diagonalize the mass matrix with a unitary transformation that maintains the kinetic part canonical
ending up with the propagating field basis $\widetilde A$. We have $A={\cal U} \widetilde A$, with
\begin{equation}
{\cal U} =\   \left( \begin{array}{ccc}
              1                              & \epsilon\, \f{m_1^2}{m_0^2-m_1^2} &
              \epsilon\, \f{m_2^2}{m_0^2-m_2^2}  \\
              \epsilon\, \f{m_0^2}{m_1^2-m_0^2} & 1                              &   0          \\
              \epsilon\, \f{m_0^2}{m_2^2-m_0^2} & 0                              &     1
 \end{array}\right)
 \label{U}
\end{equation}
We see that the interacting and the propagating photon differ by little admixtures of
$O(\epsilon)$. (We work at first order in $\epsilon$).

We have developed a quite general two paraphoton model. The specific
model we adopt has the following characteristics. First, only one
paraphoton has a mass, say $m_1\equiv \mu \neq 0$, and $m_2=0$.
Second, in order to get the effects we desire we have to assign
opposite 1 and 2 paracharges  to $f$, so that the interaction for
$f$ appearing in the last term of (\ref{L_complete}) is
\begin{equation}
e\, \bar{f} \gamma_\mu f \, ( A_1^\mu -  A_2^\mu)
 \label{Lf}
\end{equation}

\begin{figure}[b]
 \vspace{2.6cm}
 \includegraphics{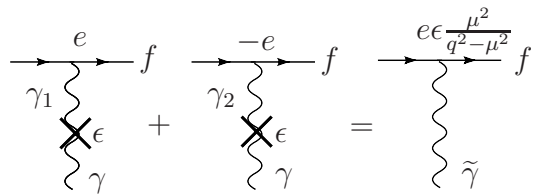}
 \caption{\it             Diagrams of the interaction of $f$ with
 photons.
   \label{fig2} }
\end{figure}

Let us show why we choose these properties. The coupling of $f$ to
photons in the interaction basis is  shown in Fig.(\ref{fig2}).
It proceeds through both paraphotons, with a relative minus sign
among the two diagrams due to the assignment (\ref{Lf}). The induced
electric $f$ charge is thus
\begin{equation}
q_f  =  {\cal U}_{10} - {\cal U}_{20}
\end{equation}

We see from (\ref{U}) that  $m_2=0$ implies  ${\cal U}_{20}=-\epsilon$. However, the value for
${\cal U}_{10}$ has to be discussed separately in the vacuum and the plasma cases. In vacuum, we
have $m_0=0$, so that ${\cal U}_{10}=0$ and thus $q_f = \epsilon$. In this case $A_0$ and $A_2$ are
degenerate and we can make arbitrary rotations in their sector. This corresponds to different
charge assignments that of course leave the physics unchanged. Due to our method of handling the
diagonalizations, eq.\eqref{U} is bad behaved for $m_0=m_2$, except for the case of our interest,
$m_0=m_2=0$, in which the order in which we take the limits $m_0\rightarrow 0$ and $m_2\rightarrow
0$ gives different charge assignments according to the rotational freedom. Here we have made
$m_2\rightarrow 0$ before $m_0\rightarrow 0$ to provide f with a milielectric charge as in
\cite{Holdom:1985ag}. Changing the order of the limits would end with a paracharge to electrons.

In the classical and non-degenerated plasmas we consider the dispersion relation can be taken as
$k^2= \omega_P^2 = 4\pi\alpha n_e/m_e $ ($n_e$ and $m_e$ are the density and mass of electrons). If
$m_0= \omega_P$ is much greater that $m_1=\mu$  we get ${\cal U}_{01}\simeq -\epsilon + \epsilon
m_1^2/m_0^2$ and the induced electric charge
\begin{equation}
q_f (k^2 \simeq \omega_P^2)\,  \simeq\,  \frac{\mu^2}{\omega_P^2} \ q_f (k^2 \simeq 0)
 \label{q_f(T)} \ \ \ .
\end{equation}

Provided we have a low energy scale $\mu \ll \omega_P\sim $ keV, we
reach our objective of having a strong decrease of the $f$ charge in
the plasma, i.e., $q_f(\omega_P^2) \ll q_f(0)= \epsilon$.

The cancelation of the two diagrams of Fig.\ref{fig2} requires that the equality $e_1=e_2$ holds up
to terms of order $O(\mu^2/\omega^2_P)$. Note that even if $e_1=e_2$ at some high energy scale
because of a symmetry, a difference in the beta functions could also spoil our mechanism at low
energy. The parafermion $f$ contributes equally to both beta functions so the problem are the
contributions from the sector that gives mass only to $A_1$. However these contributions can be
made arbitrarily small by sending the Higgs mass to infinity in the spirit of the non-linear
realizations of symmetry breaking, by considering Higgsless models like breaking the symmetry
geometrically, or by considering gauge coupling unification $e_1=e_2$ at an energy not far from the
typical solar temperature. A further possibility is to consider $e_1=e_2\ll e$ which would suppress
the loop-induced effects at the prize of making the model less natural.
\section{The role of the low-energy scale}
We now discuss the consequences of our model. The PVLAS experiment is in vacuum, so $f$ has an
effective electric charge $q_f(0)=\epsilon$, which from (\ref{triangle}) has to be
\begin{equation}
\epsilon^2 \simeq   10^{-12}\, \frac{v}{\rm eV}
 \label{N1}
\end{equation}

Concerning the astrophysical constraints, we notice that the amplitude for the Primakoff effect
$\gamma Z \rightarrow \phi Z$ is of order $q_f^2=\epsilon^2$ and that there are production
processes with amplitudes of order $\epsilon$ which will be more effective. One is plasmon decay
$\gamma^* \rightarrow \bar f f$. Energy loss arguments in horizontal-branch (HB) stars
\cite{Davidson:1991si} limits $q_f$ to be below $2\times 10^{-14}$, which translates in our model
into the bound
\begin{equation}
\epsilon\  \frac{\mu^2}{ \rm eV^2} <   4\times 10^{-8}
 \label{N2}
\end{equation}
(we have used $\omega_P \simeq 2$ keV in a typical HB core).
Other processes like bremsstrahlung of paraphotons give weaker constraints.

Equations (\ref{N1}) and  (\ref{N2}) do not fully determine the parameters of our model. Together
they imply the constraint
\begin{equation}
v\, \mu^4 <  (\, 0.4 \ {\rm eV})^5
\end{equation}
We can now make explicit one of our main results. In the reasonable
case that $v$ and $\mu$ are not too different, we wee that the new
physics scale is in the sub eV range.
\begin{figure}[b]
\includegraphics[width=9cm]{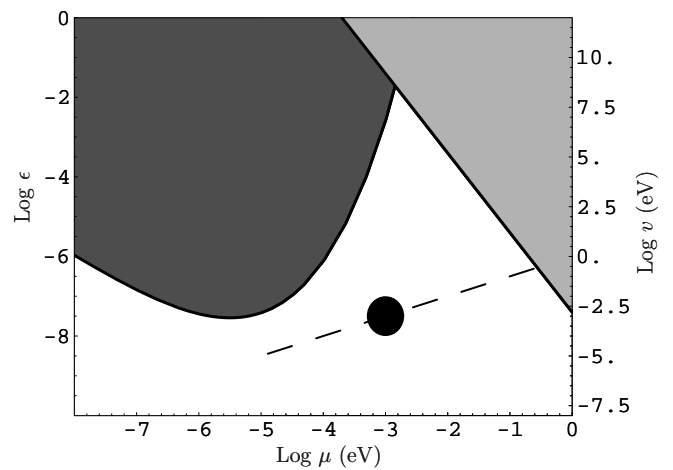}
 \caption{\it         Constraints on the parameters of our model. The black
 area is excluded  by Cavendish-type experiments,
 and the grey area by the astrophysical constraint
 (\ref{N1}). The dashed line corresponds to $v=\mu$, and the dot to
 $v=\mu\simeq 1$ meV.
   \label{fig3} }
\end{figure}

Let us consider now the CAST limit.The CAST helioscope looks for $\phi$'s with energies within a
window of 1-15 keV. In our model, $f$'s and paraphotons are emitted from the Sun, but we should
watch out $\phi$ production. This depends on the specific characteristics of $\phi$. We consider
three possibilities. A) $\phi$ is a fundamental particle. As we said the Primakoff production is
very much suppressed, so production takes place mainly through plasmon decay $\gamma^*\rightarrow
\bar{f}f \phi$. The $\phi$-flux is suppressed, but, most importantly, the average $\phi$ energy is
much less than $\omega_P \simeq.3$ keV, the solar plasmon mass. The spectrum then will be below the
present CAST energy window. B) $\phi$ is a composite $\bar{f} f$ particle confined by new strong
confining forces. The final products of plasmon decay would be a cascade of $\phi$'s and other
resonances which again would not have enough energy to be detected by CAST. C) $\phi$ is a
positronium-like bound state of $\bar{f} f$, with paraphotons providing the necessary binding
force. As the binding energy is necessarily small, ALPs are not produced in the solar plasma.

Let us now turn our attention to other constraints. Laboratory bounds on epsilon-charged particles
are much milder than the astrophysical limits, as shown in \cite{Davidson:1991si}. In our model,
however, even though paraphotons do no couple to bulk electrically neutral matter, a massive
paraphoton $\widetilde A_1$ couples to electrons with a strength $\epsilon$ and a range $\mu^{-1}$.
This potential effect is limited by Cavendish type experiments \cite{Bartlett:1988yy}.

In Fig.(\ref{fig3}) we show these limits, as well as the astrophysical bound (\ref{N2}). In the
ordinates we can see both $\epsilon$ and $v$, since we assume they are related by (\ref{N1}). At
the view of the figure, we  find out that there is wide room for the parameters of our model.
However we would like that $v$ and $\mu$ do not differ too much among them. We display the line
$v=\mu$, the region where this kind of naturality condition is fulfilled. The most economical
version of the model would be obtained when the new scales are, on the order of magnitude, about
the scale of the ALP mass of O(1 meV), (\ref{m_phi}). We have also indicated this privileged point
in the parameter space.

Also, we should discuss cosmological constraints, i.e. production of paraphotons and $f$'s in the
early universe. Taking into account that the vertices have suppression factors in the high
temperatures of such environment, we find that there is not a relic density of any of them.

Finally, let us come back to the physics responsible for the $A_1$ mass. If this comes from an
abelian Higgs mechanism then the Higgs boson acquires a milicharge $\varepsilon e_1$ and could be
produced in the Sun and in the early universe, particularly in the period of primordial
nucleosynthesis. However, this is not a problem if the mass of the Higgs is large enough, a
constraint that we required at the end of section \ref{sectiontriangle} when discussing charge
running.

\section{Conclusions}
We have presented a model of new physics containing a paracharged
particle $f$ and two paraphotons, one of which has a mass $\mu$ that
sets the low energy scale of the model. With convenient assignments
of the $f$ paracharges and mixings, we get an induced epsilon-charge
for $f$ that moreover decreases sharply in a plasma with $\omega_P
\gg \mu$. Our model accommodates an axion-like particle with the
properties (\ref{m_phi}) and (\ref{M_PVLAS}), able to explain the
PVLAS results, while at the same time consistent with the
astrophysical and the laboratory constraints, including the limit
obtained by CAST.

We have some freedom in the parameter space of our model; however if
we wish that the energy scales appearing in it are not too
different, we are led to scales in the sub eV range. A preferred
scale is O(meV), because then it is on the same order than the
axion-like particle mass.

If the interpretation of the PVLAS experiment is confirmed, which
means the exciting discovery of an axion-like particle,  then to
make it compatible with the CAST results and with the astrophysical
bounds requires further new physics. In our model, the scale of this
new physics is below the eV.

\textit{Note added}: Recently, a paper has appeared \cite{ringstring} that justifies our model in
the context of string theory.
\section{acknowledgments}
We acknowledge support by the projects FPA2005-05904 (CICYT) and
2005SGR00916 (DURSI).

\end{document}